\documentclass[fleqn,12pt]{wlscirep}
\usepackage[utf8]{inputenc}
\usepackage[T1]{fontenc}
\usepackage{setspace}
\usepackage[normal]{caption}
\usepackage{lineno}

\title{Separation of Wigner and Continuum-continuum Delays by Mirror-symmetry-broken Attosecond Interferometry}

\author[1,*]{Meng Han}
\author[1]{Jia-Bao Ji}
\author[1]{Leung Chung Sum}
\author[1,2]{Kiyoshi Ueda}
\author[1]{Hans Jakob Wörner}
\affil[1]{Laboratorium für Physikalische Chemie, ETH Zürich, Zürich, 8093, Switzerland}
\affil[2]{Department of Chemistry, Tohoku University, Sendai, 980-8578, Japan}
\affil[*]{meng.han@phys.chem.ethz.ch}

\begin{abstract}
Photoionization of matter is one of the fastest electronic processes in nature. Experimental measurements of photoionization dynamics have become possible through attosecond metrology. However, all experiments reported to date contain a so-far unavoidable measurement-induced contribution, known as continuum-continuum (CC) or Coulomb-laser-coupling delay. Exploiting the recently characterized circularly polarized attosecond pulse trains, we introduce the concept of mirror-symmetry-broken attosecond interferometry, which enables the direct and separate measurement of both the native one-photon ionization delays as well as the continuum-continuum delays. Our technique solves the longstanding challenge of experimentally isolating both the native one-photon-ionization (or Wigner) delays and the measurement-induced (CC) delays. This advance opens the door to a new generation of precision measurements that is likely to drive major progress in experimental and theoretical attosecond science with implications for benchmarking the accuracy of electronic-structure and electron-dynamics methods.
\end{abstract}

\begin{document}
\let\oldequation\equation
\let\oldendequation\endequation

\renewenvironment{equation}{\linenomathNonumbers\oldequation}{\oldendequation\endlinenomath}

%\linenumbers 
\flushbottom
\maketitle

\newpage

The birth and development of attosecond science have enabled the real-time observation, measurement and control of ultrafast electron dynamics in atoms\cite{Hentschel2001,Paul2001,jiang2022atomic}, molecules\cite{haessler2009phase,Huppert2016,heck21a,li2022attosecond,gong2022asymmetric}, clusters\cite{gong2022attosecond,Heck2022Two}, solids\cite{Ghimire2011,luu2015extreme,Lu2019,Hammond2017} and liquids\cite{Luu2018,Jordan2020}. One of the most active research fields in this area studies the question how fast an electron escapes from its binding potential in photoionization. This problem was first addressed by attosecond streaking\cite{cavalieri2007attosecond,schultze2010delay} and the reconstruction of attosecond beating by interference of two-photon transitions (RABBIT) \cite{klunder2011,tao16a,isinger17a}. Unfortunately, neither attosecond streaking nor RABBIT can directly access the delays that are truly associated with one-photon ionization. Instead, both techniques leave an indelible, measurement-induced mark on the experimentally determined delays, which is referred to as Coulomb-laser-coupling or CC delay, respectively. Importantly, this contribution is generally neither additive nor trivial to calculate \cite{baykusheva17a}. In the limiting case of a single partial wave, the contribution becomes additive\cite{dahlstrom12a,baykusheva17a}, but it still depends on the target system and the orbital angular momentum of the emitted partial wave(s), a dependence that only vanishes under the asymptotic approximation \cite{dahlstrom12a}. Moreover, the CC delays are generally similar in magnitude to the Wigner delays, such that they cannot be neglected for any quantitative purpose. The compromise adopted in most studies to date consists in measuring the difference of the Wigner delays between different species of very similar ionization energies\cite{gong2022attosecond,Heck2022Two}, atomic energy shells\cite{schultze2010delay,klunder2011} or molecular ionic states\cite{Huppert2016,heck21a}, in order to minimize the contributions of the CC delays. Adding angular resolution, the difference of the CC delays between different partial waves has also been determined \cite{fuchs2020time} and the Wigner delays have been separated into their different angular momentum components, under the assumption that the CC delays were universal and equal to the asymptotic approximation \cite{peschel2022attosecond}. However, how to experimentally separate and measure the Wigner and CC delays, is an important, yet unresolved problem in attosecond science.

\begin{figure}[htbp]
\centering
\includegraphics[width=11.5cm]{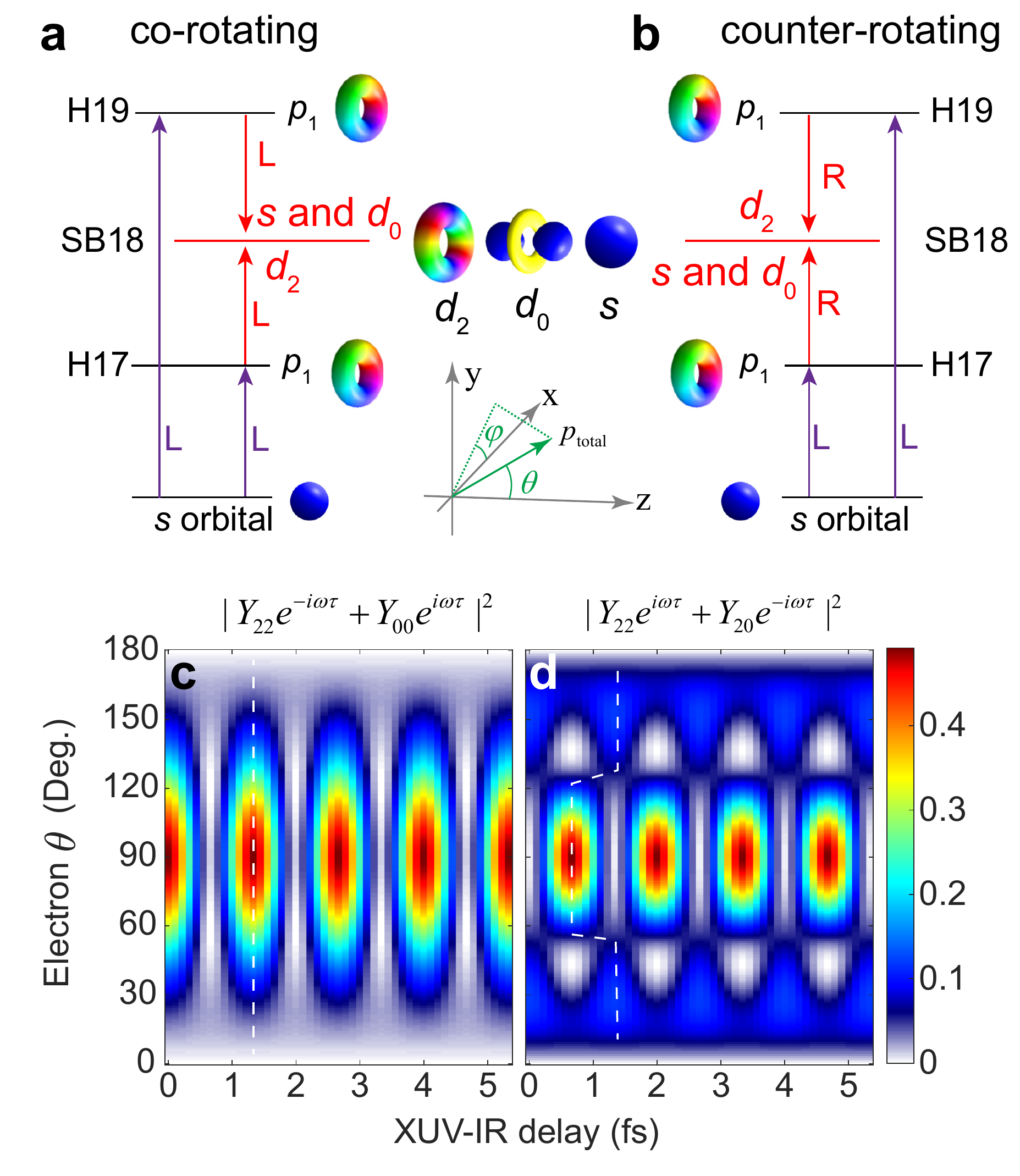}
\caption{\textbf{Measurement principle of mirror-symmetry-broken attosecond interferometry}. \textbf{a, b}, Partial-wave diagrams in co-rotating and counter-rotating bi-circularly polarized XUV-IR fields, respectively. In our coordinate system, the $z$ axis (i.e., quantum axis of atomic orbitals) is defined as the light propagation direction and the $x-y$ plane is the light polarization plane. In both geometries, the electron wavepackets at sidebands is a coherent superposition of $d_2$, $d_0$ and $s$ partial waves, but with different amplitudes and phases, as a result of mirror symmetry breaking. \textbf{c, d}, Calculated two-dimensional interference patterns between $d_2$ and $s$ or $d_0$ waves with equal amplitudes and phases, respectively. The constant phase fronts, i.e., angle-resolved RABBIT phases, are added in c and d with dashed lines.}
\label{fig:figure1}
\end{figure}

Here, we solve this challenge by moving away from linear polarizations, employing circularly polarized XUV and IR fields\cite{Han2022ACDC}, and introducing the concept of mirror-symmetry-broken attosecond interferometry. A left circularly polarized attosecond pulse train generated from the non-collinear high-order harmonic generation (HHG) process\cite{Hickstein2015,Huang2018,han2022attosecond} of argon is used to photoionize helium atoms, preparing an electronic $p_{1}$ continuum state, where $p$ represents the orbital angular momentum quantum number and the subscript is the magnetic quantum number. Note that here the quantization axis is defined as the light propagation direction ($z$ axis) due to the circular polarization, as illustrated in Fig.~\ref{fig:figure1}. The $p_{1}$ orbital carries the spiral phase front with the same helicity as the XUV field. A co-rotating or counter-rotating circularly polarized IR field is used to probe the Wigner phase of the photoelectron wave packet through the RABBIT technique\cite{Paul2001,klunder2011,dahlstrom2013theory}. Three-dimensional momenta of photoelectrons and photoions were measured in coincidence with a COLTRIMS spectrometer\cite{Dorner2000,Ullrich2003}. Experimental details are given in the Methods section. The sideband generated by the XUV-IR two-photon transition is a coherent superposition of $d_{2}$, $d_{0}$ and $s$ partial waves in both co- and counter-rotating geometries. In the co-rotating geometry, the absorption pathway gives rise to the $d_{2}$ wave and the emission pathway creates the mixture of $d_{0}$ and $s$ waves, while the partial waves become mirror-reversed in the counter-rotating geometry. This mirror symmetry between co-rotating and counter-rotating geometries will be broken if one is able to resolve the partial-wave amplitudes and phases contributing to the sideband. This is due to the helical or chiral phase front of the XUV-induced electron vortices\cite{djiokap2015electron} at the main peaks, which will trigger a dichroic response to a circularly polarized IR probe. The symmetry breaking results in circular dichroism (CD) on the partial-wave amplitudes and phases. The different partial-wave amplitudes ($A_{d_2}^{\rm{co/counter}},A_{d_0}^{\rm{co/counter}},A_{s}^{\rm{co/counter}}$) and phases ($\phi_{d_2}^{\rm{co/counter}},\phi_{d_0}^{\rm{co/counter}},\phi_{s}^{\rm{co/counter}}$) will give rise to the significantly different interference patterns for the sideband electrons as a function of the XUV-IR delay $\tau$ and the photoelectron emission angle $\theta$ with respect to the light propagation direction, i.e., $I^{\rm{co}}(\theta,\tau) = |A_{d_2}^{\rm{co}}Y_{22}(\theta)e^{i\omega \tau + i\phi_{d_2}^{\rm{co}}} + A_{d_0}^{\rm{co}}Y_{20}(\theta)e^{-i\omega \tau + i\phi_{d_0}^{\rm{co}}} + A_{s}^{\rm{co}}Y_{00}(\theta)e^{-i\omega \tau + i\phi_{s}^{\rm{co}}}|^2$ and $I^{\rm{counter}}(\theta,\tau) = |A_{d_2}^{\rm{counter}}Y_{22}(\theta)e^{-i\omega \tau + i\phi_{d_2}^{\rm{counter}}} + A_{d_0}^{\rm{counter}}Y_{20}(\theta)e^{i\omega \tau + i\phi_{d_0}^{\rm{counter}}} + A_{s}^{\rm{counter}}Y_{00}(\theta)e^{i\omega \tau + i\phi_{s}^{\rm{counter}}}|^2$, where $Y_{lm}$ are the spherical harmonic functions. Note that the interference pattern doesn't depend on the photoelectron $\varphi$ angle in the polarization plane because of the circularity of both fields. By fitting the two-dimensional interference pattern globally\cite{villeneuve2017coherent} with above formulas, one can extract the partial-wave amplitudes and phases in the two geometries, respectively. The Wigner phase and CC phase can then be obtained separately by simple mathematical operations on these extracted partial-wave phases.

The partial-wave phase on the sideband is essentially the phase of the two-photon transition matrix element\cite{dahlstrom2013theory}, and in the perturbative regime it can be further decoupled into the sum of $-l\pi/2$ (centrifugal phase factor of the orbital angular momentum $l$), $\phi_{s\rightarrow p_1}^{\rm{Wigner}}(E_k\pm \omega)$ (Wigner phase of single-photon ionization from $s$ state to $p_1$ state) and $\phi_{\rm{absorption/emission}}^{\rm{CC}~(\textit{p}_1\rightarrow \textit{l}_\textit{m})}(E_k\pm\omega;E_k)$ (CC phase from $p_1$ state to $l_m$ state), where $E_k$ is the sideband energy and $\omega$ is the IR photon energy. When the electron energy is away from the threshold, the CC phases for absorption and emission pathways are opposite with each other\cite{dahlstrom2013theory}, i.e., $\phi_{\rm{absorption}}^{\rm{CC}}(E_k-\omega;E_k) = -\phi_{\rm{emission}}^{\rm{CC}}(E_k+\omega;E_k)$. Therefore, the phase sum for the same partial wave in the two geometries cancels out the CC phase and gives rise to the Wigner phase solely. On the other hand, the phase difference between the two geometries will cancel out most of the Wigner phase, isolating the CC phase approximately. For example, the phase sum and the phase difference for the $d_2$ wave can be described by 
\begin{equation}
\begin{aligned}
    \phi_{d_2}^{\rm{co}} +\phi_{d_2}^{\rm{counter}} &= [\phi_{s\rightarrow p_1}^{\rm{Wigner}}(E_k-\omega) + \phi_{s\rightarrow p_1}^{\rm{Wigner}}(E_k+\omega) ] + [\phi_{\rm{absorption}}^{\rm{CC}~(\textit{p}_1\rightarrow \textit{d}_2)}(E_k-\omega;E_k) + \phi_{\rm{emission}}^{\rm{CC}~(\textit{p}_1\rightarrow \textit{d}_2)}(E_k+\omega;E_k) ]\\
    &\overset{E_k>>\omega}{=}2\phi_{s\rightarrow p_1}^{\rm{Wigner}}(E_k) 
\end{aligned}    
\end{equation}
and
\begin{equation}
\begin{aligned}
    \phi_{d_2}^{\rm{co}} -\phi_{d_2}^{\rm{counter}} &= [\phi_{s\rightarrow p_1}^{\rm{Wigner}}(E_k-\omega) - \phi_{s\rightarrow p_1}^{\rm{Wigner}}(E_k+\omega) ] + [\phi_{\rm{absorption}}^{\rm{CC}~(\textit{p}_1\rightarrow \textit{d}_2)}(E_k-\omega;E_k) - \phi_{\rm{emission}}^{\rm{CC}~(\textit{p}_1\rightarrow \textit{d}_2)}(E_k+\omega;E_k) ]\\
    &\overset{E_k>>\omega}{=}2\omega\tau_{s\rightarrow \textit{p}_1}^{\rm{Wigner}}(E_k) + 2\phi_{\rm{absorption}}^{\rm{CC}~(\textit{p}_1\rightarrow \textit{d}_2)}\;,
\end{aligned}    
\end{equation}
respectively. Using the discrete Wigner phases obtained through Eq. (1), one can determine the Wigner delay by the finite differential $\tau_{s\rightarrow p_1}^{\rm{Wigner}} \approx \frac{\Delta\phi_{s\rightarrow p_1}^{\rm{Wigner}}(E_k)}{\Delta E_k}$. Then one can obtain the CC phase by substituting $\tau_{s\rightarrow p_1}^{\rm{Wigner}}$ in Eq. (2) and also the CC delay by the finite differential of the resulting CC phase. As for the $d_0$ wave, the phase sum gives the same result as the $d_2$ wave, i.e., the Wigner phase of the $1s^2$ to $1s\epsilon p$ transition, while the phase difference defined by the co-rotating result subtracting the counter-rotating one, gives the CC phase in the emission pathway. 
For the $s$ wave, the phase sum will show an additional $\pi$ shift compared with the results from the $d_2$ and $d_0$ channels due to the contribution of the centrifugal phase factor of $-l\pi/2$. 

We first display the basic structure of interference patterns in the two geometries. Due to Fano's propensity rule\cite{fano1985propensity,Busto2019Fano}, i.e., absorbing (emitting) a photon preferentially changes $l$ by +1 (-1). The sideband in the co-rotating case should be more dominated by the interference between $d_2$ and $s$ waves. In contrast, the sideband in the counter-rotating case will be more dominated by the interference between $d_2$ and $d_0$ waves. In Figs. \ref{fig:figure1}c and d, we show the interference pattern between $d_2$ and $s$ and that between $d_2$ and $d_0$, respectively, with the assumption of equal partial-wave phases and amplitudes. For the co-rotating geometry, the interference between $d_2$ and $s$ displays the isotropic phase profile, which indicates the RABBIT phase is angle independent. For the counter-rotating geometry, the interference between $d_2$ and $d_0$ shows the $\pi$-shift structure between the light polarization plane $\theta=90^\circ$ and the light propagation direction $\theta=0^\circ$. The $\pi$-shift originates from the alternating sign of the three lobes of the $d_0$ wave. Hence, the two-dimensional interference pattern is very sensitive to the amplitudes and phases of participated partial waves, allowing to accurately retrieve these parameters.

\begin{figure}[htbp]
\centering
\includegraphics[width=12.0cm]{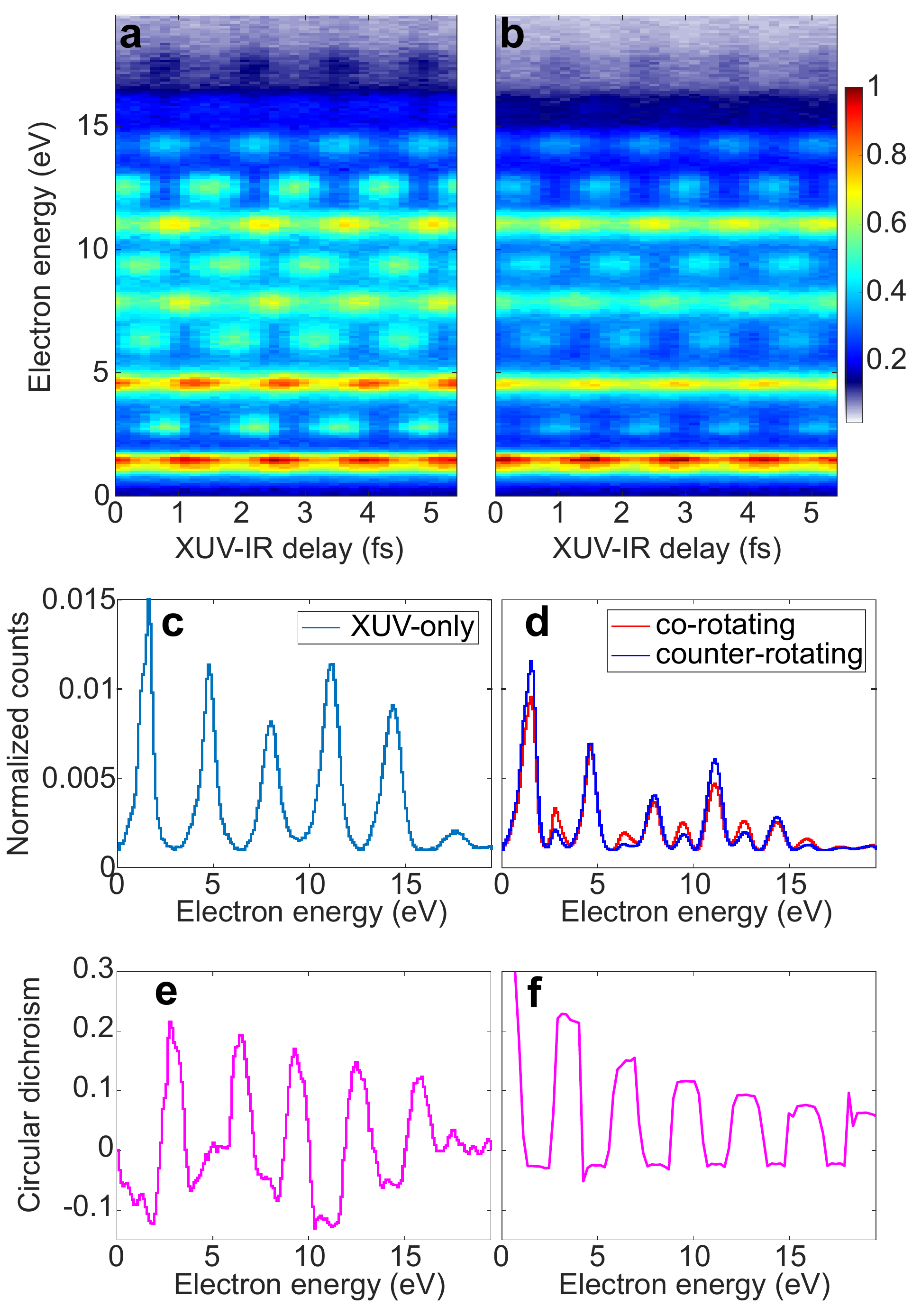}
\caption{\textbf{Circular dichroism on the yield of sideband electrons}. \textbf{a, b}, Measured photoelectron RABBIT traces in co-rotating and counter-rotating geometries, respectively. \textbf{c, d}, Measured photoelectron energy spectra in the XUV-only and two-color fields, respectively. \textbf{e, f}, Measured and simulated photoelectron circular dichroism between co-rotating and counter-rotating geometries, respectively. Because of the circularity of both fields, the RABBIT traces and the CD effect don't depend on the photoelectron $\varphi$ angle. Here, in both experiment and simulation the photoelectron $\varphi$ angle is fixed at zero with a open angle of $20^\circ$ and the $\theta$ angle is integrated over its $\pi$ range., where $\varphi$ and $\theta$ angles are defined in Fig. \ref{fig:figure1}. }
\label{fig:figure2}
\end{figure}

In Fig. \ref{fig:figure2}, we show the CD effect on the sideband yield, where the panels a and b are the measured delay-resolved $\theta$-averaged photoelectron energy spectra (i.e., RABBIT traces) in the co-rotating and counter-rotating geometries, respectively. The two RABBIT traces were recorded under the same conditions, including the same starting point of the scan, except for the IR helicity. The XUV-only energy spectrum is shown in panel c and the delay-averaged energy spectra in the bi-circular fields are illustrated in panel d, where each spectrum was normalized to its own count sum. Obviously, the sideband signal in the co-rotating geometry is much stronger than in the counter-rotating geometry. To quantify the CD effect, we introduce an energy-dependent CD defined as $[Y_{\textbf{co}}(E_k) - Y_{\textbf{counter}}(E_k)]/[Y_{\textbf{co}}(E_k) +Y_{\textbf{counter}}(E_k)]$, where $Y_{\textbf{co}}(E_k)$ and $Y_{\textbf{counter}}(E_k)$ are the normalized photoelectron counts in the co- and counter-rotating cases, respectively. Our measured CD (Fig. \ref{fig:figure2}e) can reach up to 20$\%$ around 3 eV (i.e., SB18) and then it gradually decreases when the electron energy goes up. Our findings are supported by the solutions of the time-dependent Schr\"odinger equation (TDSE) in three dimensions (see Fig. \ref{fig:figure2}f). Simulation details are given in the Methods section.

\begin{figure}[htbp]
\centering
\includegraphics[width=10.0cm]{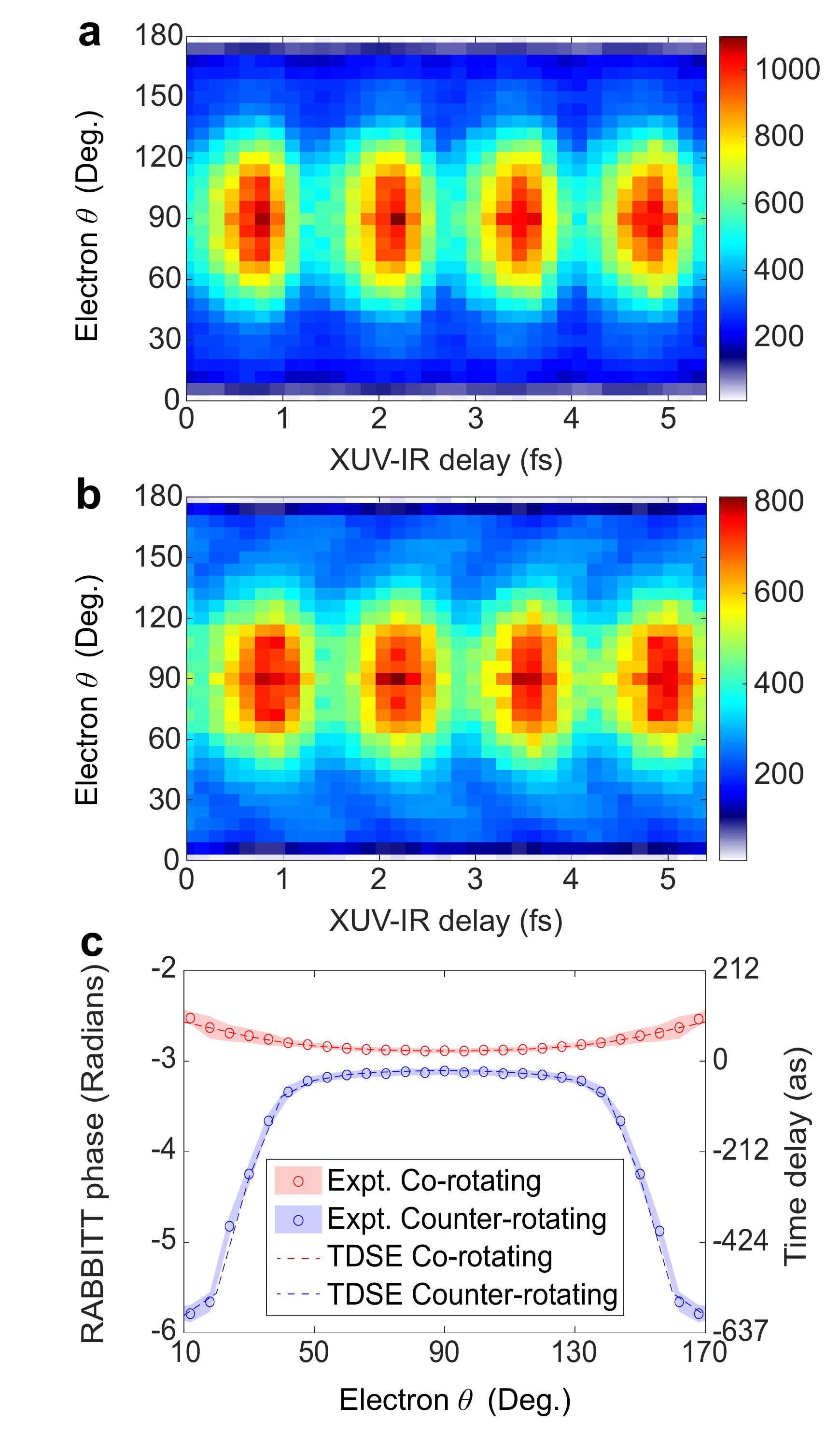}
\caption{\textbf{Circular dichroism on the RABBIT phase of sideband electrons}. \textbf{a, b}, Measured $\theta$-resolved RABBIT traces of SB18 in co-rotating and counter-rotating geometries, respectively. \textbf{c}, Extracted $\theta$-resolved RABBIT phases in the two geometries. The uncertainty (shaded area) is estimated by the backgroupd-over-amplitude approach\cite{Ji2021}.}
\label{fig:figure3}
\end{figure}

We next discuss the CD effect on the RABBIT phase. In Figs. \ref{fig:figure3}a-b, we display the measured two-dimensional interference patterns of SB18 in the co-rotating and counter-rotating geometries, respectively. Comparing with the simplified model of two-wave interference shown in Figs. \ref{fig:figure1}c-d, here, the experimental results reveal more rich and detailed structures. At each emission angle, we perform Fourier transformation to extract the phase of the yield oscillation, i.e. the so-called angle-resolved RABBIT phase, which is shown in Fig. \ref{fig:figure3}c. There is not only a phase gap between the two geometries, but also the RABBIT phase curves have the opposite curvature in the two geometries, and thus the phase gap between them is angle dependent. In the light-polarization plane (i.e. $\theta = 90^\circ$), the converted photoionization time delay difference is about 55 attoseconds and it increases to around 700 attoseconds in the light-propagation direction. The $\pi$-shift structure between the polarization plane and the light-propagation direction in the counter-rotating geometry is still visible but more smooth, since here three partial waves participate in the interference. Our high-resolution experimental results allow for a quantitative comparison with the TDSE results, as illustrated with dashed lines in Fig. \ref{fig:figure3}c. The TDSE simulations agree excellently with our measurements, validating the breaking of mirror symmetry and the induced CD effect on the two-photon ionization time delay.

\begin{figure}[htbp]
\centering
\includegraphics[width=14.5cm]{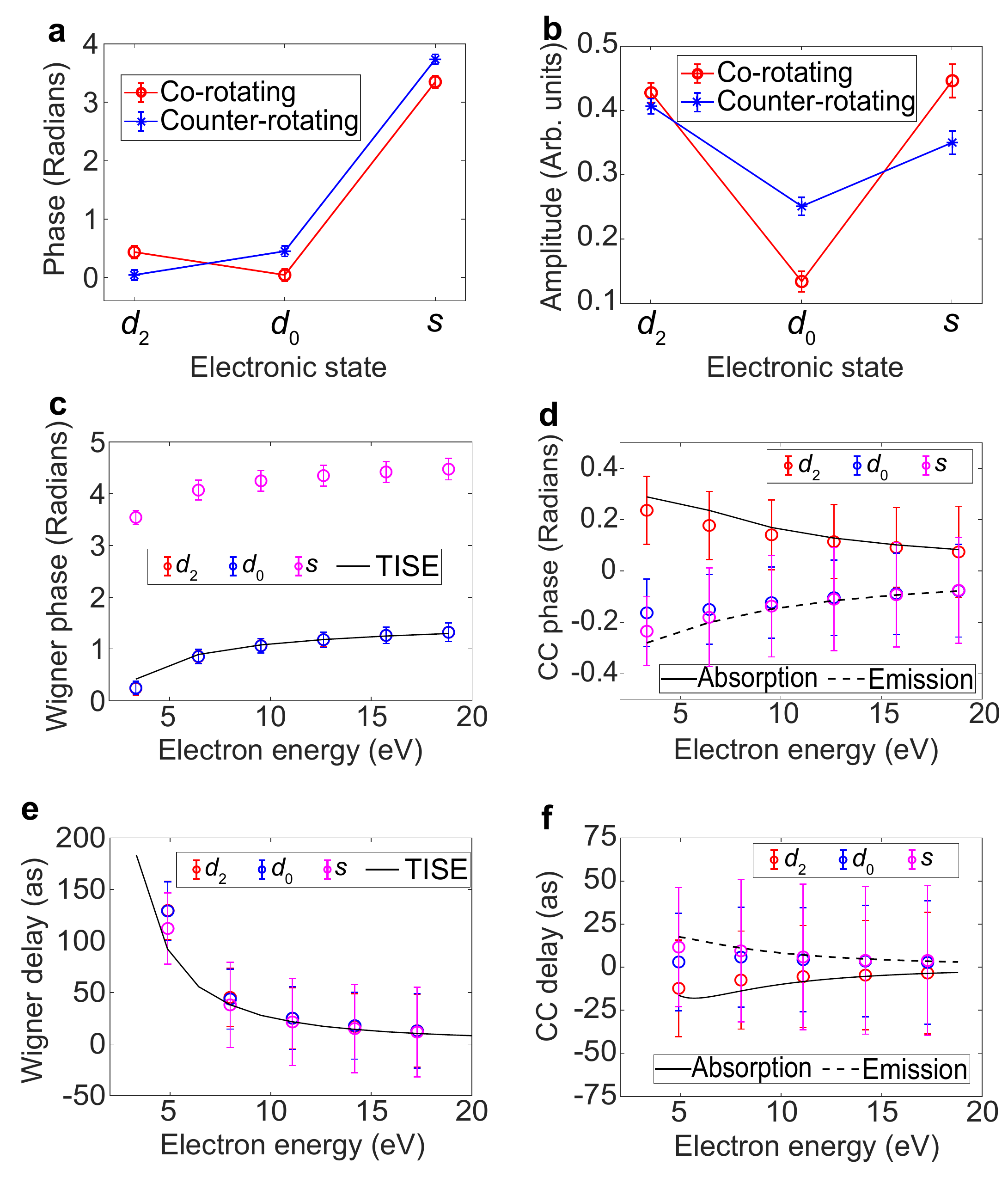}
\caption{\textbf{Separation of Wigner and CC delays}. \textbf{a, b}, Experimentally retrieved amplitudes and phases of the relevant partial waves for SB 18, respectively. The error bars represent the standard fitting errors within the 95\% confidence interval. \textbf{c, d}, Experimentally separated Wigner and CC phases, respectively. In c, the black solid line represents the Wigner phase of helium $1s^2$ to $1s\epsilon p$ transition calculated by the time-independent Schr\"odinger equation. In d, the absorption and emission CC phases are the predictions by the analytical formula based on the long-range-corrected asymptotic wave functions of hydrogen (i.e., Eq. 30 in the reference\cite{dahlstrom2013theory}).\textbf{e, f}, Wigner and CC delays obtained by the finite differential from the result in c and d. In c-f, the error bars are determined by the error propagation formula with the uncertainties of the retrieved partial-wave phases.}
\label{fig:figure4}
\end{figure}

Based on the two-dimensional interference patterns, we can extract the partial-wave amplitudes and phases by global fitting\cite{villeneuve2017coherent,Han2022ACDC} with the interference formulas. The retrieved partial-wave amplitudes and phases of SB 18 as shown in Figs. \ref{fig:figure4}a-b. For more details on the data analysis, see the Method section and Extended-data figures. The retrieved partial-wave phases show a notable dichroic feature between the two geometries: the phase increases monotonically from $d_2$ over $d_0$ to $s$ states in the counter-rotating geometry, while there is a phase minimum at the $d_0$ state in the co-rotating geometry. As for the partial-wave amplitudes, there is also a significant dichroic feature for the relative magnitudes between $d_0$ and $s$ waves in the two geometries. After we retrieved the partial-wave phases in the two geometries, we can obtain the Wigner phase and the CC phase using Eqs. (1-2), which are illustrated in Figs. \ref{fig:figure4}c-d, respectively. For the Wigner phase, the results from $d_2$ and $d_0$ are almost identical and there is a constant phase difference of $\pi$  between the results from $s$ and $d_{2,1}$ due to the different centrifugal phase factor. We compare our experimental results with the calculated Wigner phase of helium $1s^2$ to $1s\epsilon p$ transition by solving the time-independent Schr\"odinger equation (TISE) using the single-electron approximation. Our experimental results agrees very well with the TISE results when the electron energy is larger than 7 eV, as shown in Fig. \ref{fig:figure4}c. In Fig. \ref{fig:figure4}d, we illustrate the retrieved CC phases from the three partial waves. The $d_2$ wave gives rise the CC phase of absorbing one IR photon to the SB positions, while the $d_0$ and $s$ waves result in the CC phase in the emission channel, allowing to access the $l$-dependence of the CC phase\cite{fuchs2020time}. We compare the retrieved CC phases with the prediction by the analytical formula based on asymptotic wave functions of the hydrogen atom without the $l$-dependence (i.e., Eq. 30 in the reference\cite{dahlstrom2013theory}), as illustrated in Fig. \ref{fig:figure4}d. In our experimental results, the $l$-dependent feature can be observed on the low-order sidebands \cite{boll2022analytical}. When the electron energy is larger, our retrieved CC phases are more closer to the analytical predictions due to the assumption we used in Eqs. (1-2). In Fig. \ref{fig:figure4}e-f, we evaluate the Wigner and CC delays using the corresponding phases by finite numerical differential to approximate the energy derivative. With such a simple and transparent retrieval method, our approach has already enabled to quantitatively retrieve the Wigner and CC delays, which can directly and separately be compared to theoretical calculations.

In summary, we have demonstrated a general experimental and conceptual protocol to separately measure both the Wigner and CC delays. This separation was made possible through the innovative application of circularly polarized attosecond and infrared pulses. The mirror symmetry between the co-rotating and counter-rotating geometries is broken due to the spiral phase front of the photoelectron vortices created by circularly polarized XUV attosecond pulses, which gives rise to the significant CD on both the amplitude and phase of the sideband electrons. The phase CD provides the opportunity to separate the Wigner and CC phases experimentally. Our approach has important implications for precision measurements of photoionization delays and CC delays. The availability of such precision results will motivate advances in electronic-structure and electron-dynamics methods. The current state of the art in theoretical methods indeed still largely relies on the asymptotic approximation of the CC delays \cite{dahlstrom12a,baykusheva17a,peschel2022attosecond}. This approximation has the advantage that it is convenient because it removes the target and angular-momentum dependence of the CC delays and additionally makes them separable from the one-photon delays. The application of our method to the s-shells of other atoms will enable the first quantitative studies of the target- and angular-momentum dependence of the CC delays across the periodic table. Our new methodology will also drive major progress in molecular attosecond chronoscopy. For linear molecules, the Wigner delay caused by the vortex-electron scattering with neighboring atoms can be measured if the molecule is aligned along the light propagation direction\cite{ota2021theory_a,ota2021theory_b}. For ring-shaped molecules, our approach can measure not only the amplitude but also the phase of the periodically varying electron ring current caused by the Jahn-Teller distortions\cite{nandipati2021dynamical}. Finally, our method can also be extended to probe magnetic and topological effects in solids on the attosecond time scale by detecting CD effects in amplitude and phase of photoelectron vortex scattering.

\section*{Methods}
\textbf{Experimental details.} The description of our experimental setup can be found in the references\cite{Han2022ACDC,han2022attosecond}. The near-infrared laser pulse (2 mJ) was delivered from a regenerative Ti:sapphire laser amplifier at the central wavelength of 799 nm with a repetition rate of 5 kHz. The pulse duration was compressed to be around 31 fs for the full width of half maximum measured by a home-made transient-grating FROG. This laser beam was split with a 70:30 beam splitter, and the more intense part was sent through our beam-in-beam module\cite{han2022attosecond} and then focused by a lens (f = 30 cm) onto a 3 mm long, argon-filled gas cell to generate the circularly polarized extreme-ultraviolet attosecond pulse train via non-collinear HHG. The left circularly polarized one of the two dominant XUV beams was picked up using a perforated mirror and then focused into the main chamber of the Cold Target Recoil Ion Momentum Spectrometer (COLTRIMS) by a nickel-coated toroidal mirror (f = 50 cm). The XUV spectrum was characterized with a home-built XUV spectrometer consisting of an aberration-corrected flat-field grating (Shimadzu 1200 lines/mm) and a micro-channel-plate (MCP) detector coupled to a phosphor screen. The beam with $30\%$ energy was used as the dressing field in the RABBIT experiments. The dressing IR field was adjusted to left or right circular polarization by a zero-order quarter waveplate and its intensity was controlled at a very low level (about $10^{12}$ W/cm$^2$) by an iris. The dressing IR beam was recombined with the XUV beam by the perforated mirror and then was focused by a perforated lens (f = 50 cm). In the arm of the dressing field, there were two delay stages, i.e., a high precision direct-current motor (PI, resolution 100 nm) and a piezoelectric motor (PI, resolution 0.1 nm), operating on femtosecond and attosecond time scales, respectively. A cw HeNe laser beam was sent through the beam splitter to monitor the relative lengths of the IR paths in the two arms of the interferometer. A fast CCD camera behind the perforated recombination mirror was used to image the interference fringes from the HeNe laser in order to lock the phase delay between XUV and IR by the PID feedback. When scanning the XUV-IR phase delay in the measurements, the piezo delay stage was actively stabilized with a step size of 156 as and a jitter of less than 30 as. Note that the phase lock was not stopped when switching the IR helicity in order to make sure the two measurements have the same starting point for the PI stage. For the COLTRIMS setup, the supersonic gas jet of helium atoms (backing pressure at 3 bar) was delivered along the $x$ direction by a small nozzle with a opening hole diameter of 50 $\mu m$ and passed through two conical skimmers (Beam Dynamics) located 10 mm and 30 mm downstream with a diameter of 0.2 mm and 1 mm, respectively. For the COLTRIMS spectrometer, static electric ($\sim$ 1.605 V/cm) and magnetic ($\sim$ 7.090 G) fields were applied along the $y$ axis to collect the charged fragments in coincidence. Only the single-ionization events (one electron is coincident with one He$^+$) were presented in this work.

\noindent\textbf{TDSE simulations.} We performed the TDSE simulations based on an open-source TDSE solver, Qprop 2.0\cite{Mosert2016}, where the details of the algorithm and the source code are available. In the simulation, we used the Tong-Lin model potential\cite{tong05a} $V_{\text{eff}} = -[Z_c+a_1\rm{exp}(-a_2\emph{r})+\emph{a}_3 \emph{r} \rm{exp}(-a_4\emph{r}) + \emph{a}_5\rm{exp}(-a_6\emph{r})]/\emph{r}$ for helium atoms, where $Z_c = 1$, $a_1 = 1.231$, $a_2 = 0.662$, $a_3 = -1.325$, $a_4 = 1.236$, $a_5 = -0.231$ and $a_6 = 0.480$. The vector potential of the XUV field is given by $A_\text{XUV}(t) = -A_{\text{XUV}}^{0}\sum_{i=17,19,...,29}\text{sin}^2(\omega t/2n_c)*[\text{sin}(i\omega t)\vec{x} +\text{cos}(i\omega t)\vec{y}]$ and that of the IR field is $A_\text{IR}(t) = -A_{\text{IR}}^{0} \text{sin}^2(\omega t/2n_c)*[\text{sin}(\omega (t +\tau))\vec{x}+\text{cos}(\omega (t+\tau))\vec{y}]$, where the amplitude $A_{\text{XUV}}^{0} = 0.00534~\text{a.u.}$, $A_{\text{IR}}^{0} = 0.0025~\text{a.u.}$, the light duration amounts to $n_c = 6$ cycles, and the XUV-IR delay $\tau$ was uniformly sampled by 24 points in one IR cycle. In the simulations, the discretization box of the radial part is 160 a.u. with the grid size of 0.025 a.u., and the maximum angular momentum included is $l_{\text{max}} = 10$, which are both big enough to cover all ionized electronic partial waves. The time step was $\Delta t = 0.01$ a.u.. The convergence of the numerical calculations has been checked with respect to all discretization parameters. For the Wigner phase shown in Fig. 4c, we calculated the $p$-partial-wave phase of the continuum eigenfunctions in the time-independent Schr{\"o}dinger equation.

\begin{figure}[htbp]
\centering
\includegraphics[width=9.5cm]{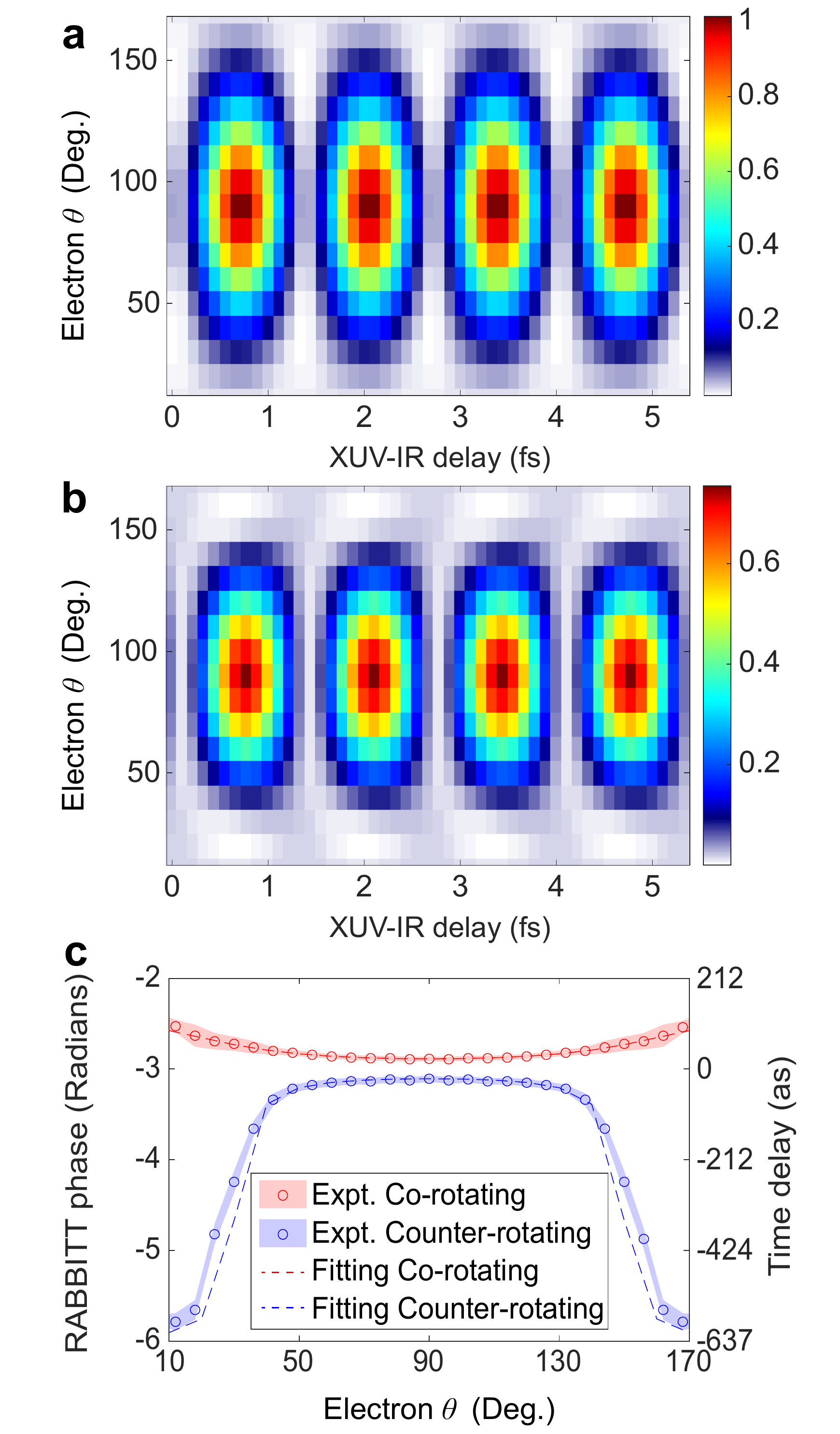}
\caption{\textbf{Global fitting}. \textbf{a, b},  Global fitting results of the experimental $\theta$-resolved RABBITT traces of SB18 in co-rotating and counter-rotating geometries shown in Figs. 3a-b of the main text, respectively. \textbf{c}, Comparison of the $\theta$-resolved RABBITT phases between experiment and global fitting.}
\label{fig:global_fitting}
\end{figure}

\noindent\textbf{Retrieval of partial-wave phases and amplitudes by global fitting.} In Fig. 3, we showed the measured two-dimensional interference patterns as a function of the XUV-IR delay $\tau$ and the emission angle $\theta$ in the co-rotating and counter-rotating geometries. The two-dimensional interference pattern of the sideband photoelectron can be described by the three-wave interference models, i.e.
\begin{equation}
\begin{aligned}
    I^{\rm{co}}(\theta,\tau) = |A_{d_2}^{\rm{co}}Y_{22}(\theta)e^{i\omega \tau + i\phi_{d_2}^{\rm{co}}} + A_{d_0}^{\rm{co}}Y_{20}(\theta)e^{-i\omega \tau + i\phi_{d_0}^{\rm{co}}} + A_{s}^{\rm{co}}Y_{00}(\theta)e^{-i\omega \tau + i\phi_{s}^{\rm{co}}}|^2 
\end{aligned}    
\end{equation}
in the co-rotating geometry and
\begin{equation}
\begin{aligned}
    I^{\rm{counter}}(\theta,\tau) = |A_{d_2}^{\rm{counter}}Y_{22}(\theta)e^{-i\omega \tau + i\phi_{d_2}^{\rm{counter}}} + A_{d_0}^{\rm{counter}}Y_{20}(\theta)e^{i\omega \tau + i\phi_{d_0}^{\rm{counter}}} + A_{s}^{\rm{counter}}Y_{00}(\theta)e^{i\omega \tau + i\phi_{s}^{\rm{counter}}}|^2 
\end{aligned}    
\end{equation}
in the counter-rotating geometry. The experimental patterns were fit to Eqs. (3-4) to obtain the partial-wave amplitudes and phases. The experimental patterns were subtracted from the model image point-by-point, and the difference-squared was summed. This resulted in a goodness-of-fit parameter in a least-squares sense. The initial values for the parameters in Eqs. (1-2) were varied to create a global optimum fit of the parameters. Note that the two patterns are fit simultaneously in order to ensure they share the same unknown phase constant. For different order sidebands, this phase constant was calibrated to the TDSE results in order to remove the effect of the XUV attochirp. In Figs. \ref{fig:global_fitting}a-b, we show the fitted patterns corresponding to the experimental patterns shown in Fig. 3. And we also compare the fitted $\theta$-resolved RABBITT phases with the experimental results in Fig. \ref{fig:global_fitting}c. Our global fitting results show the quantitative agreement with the measurement, which indicates the good accuracy of our retrieved parameters.

\section*{Data availability}
The data that support the plots within this paper and other findings of this study are available from the corresponding author upon reasonable request. Correspondence and requests for materials should be addressed to M.H..

\bibliography{pop_references,attobib}

\section*{Acknowledgements}
M. H. acknowledges the funding from the European Union’s Horizon 2020 research and innovation program under the Marie Skłodowska-Curie grant agreement No 801459 - FP-RESOMUS. This work was supported by ETH Z\"urich and the Swiss National Science Foundation through projects 200021\_172946 and the NCCR-MUST. M. H. thanks Dr. Hao Liang for the fruitful discussions on the calculation of TISE.

\section*{Author contributions}
M.H. performed the experiments with the support of J.J. and L.C.S.. M.H. analyzed and interpreted the data. Simulations were implemented by M.H.. M.H. and K.U. conceived the study and H.J.W. supervised its realization. M.H., K.U. and H.J.W. discussed the results and wrote the paper with the input of all co-authors.

\section*{Competing interests}
All co-authors have seen and agree with the contents of the manuscript and there is no financial interest to report.

\newpage
\section*{Extended-data Figures}

\begin{figure}[h!]
\centering
\includegraphics[width=16.2cm]{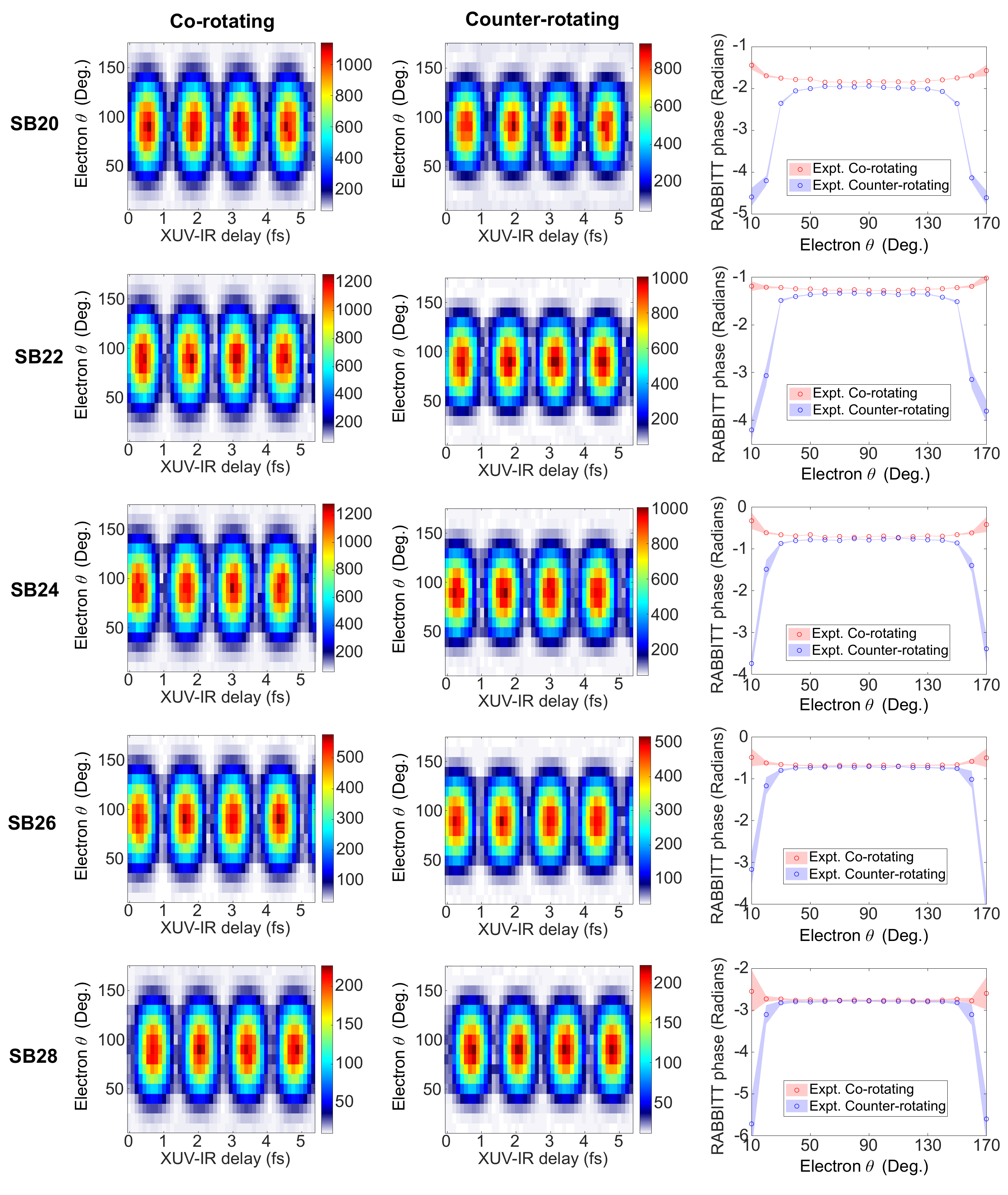}
\caption{\textbf{Experimental results from SB20 to SB28}. In Fig. 3 of main text, we show the experimental result of SB18, which has the largest CD effect. Here we supplement the results for the other sidebands. The first and second columns shows the measured RABBITT traces in co-rotating and counter-rotating geometries, respective. The third column displays the extracted RABBITT phases. From the top row to the bottom tow, they are corresponding to the results from SB20 to SB28.}
\label{fig:expt_data}
\end{figure}

\end{document}